\documentstyle[multicol,prl,aps,epsf]{revtex}
\begin{document}
\draft
\title{Cooperative Dynamics in 
Unentangled Polymer Fluids} 
\author{M. Guenza}
\address{
Institute of Theoretical Science and Material Science Institute,
University of Oregon, Eugene, OR 97403}
\maketitle
\vspace{+3mm}
\begin{abstract}

We present a Generalized Langevin Equation for the dynamics
of interacting semiflexible polymer chains, undergoing slow cooperative
dynamics.
The calculated Gaussian intermolecular center-of-mass and monomer 
potentials, wich enter the GLE,
are in quantitative agreement 
with computer simulation data. The experimentally observed, short-time 
subdiffusive regime of the polymer
mean-square displacements, emerges here from the 
competition between the intramolecular and
the intermolecular mean-force potentials. 
 
\end{abstract}  

\vspace{+3mm}

\begin{multicols}{2}
\narrowtext
Despite its intrinsic complexity, the dynamics of low molecular weight 
unentangled polymer 
fluids has been considered for a long time to be a well understood problem in 
polymer physics. 
It is commonly accepted that when the 
degree of polymerization, $N$, does not exceed the entanglement value, $N_e$, 
a molecule 
is free to diffuse in the liquid, and follows Fickian dynamics. In this case, 
the Rouse 
approach successfully describes several key features of the polymer 
dynamics. These include the scaling with $N$ of the bulk viscosity and the 
single-chain 
diffusion coefficient $D_{Rouse}=k_BT (\zeta N)^{-1}$, with $k_B$ the 
Boltzmann constant, $T$ the temperature,
and $\zeta$ the monomer friction coefficient\cite{DoiEdw}. 

Upon closer examination, however, the dynamics of unentangled polymer 
fluids still presents important unresolved questions.
In the  Rouse theory, 
local intermolecular
interactions are completely neglected; the single chain dynamics is driven by
intramolecular entropic restoring forces and segmental friction, while the 
dynamics of the surrounding chains provide a heat bath.
In this description the polymer center of mass is free to diffuse following
Brownian dynamics, and the mean-square displacement is 
expected to scale linearly in time at all time scales.
In reality, a single polymer in a fluid spans a volume $V \propto R_g^3$, with 
$R_g=\sqrt{N} l/\sqrt{6}$ the polymer radius of gyration, and $l$ the 
statistical 
segment length. Inside this volume are contained an average of
$n \propto \sqrt{N}$ chains, 
that interact with each-other through the potential of mean force\cite{hansen}.
The range of the potential is the same as that 
of the correlation 
hole\cite{DeGennes} in the pair correlation function $g({\bf r})$,
a distance of order $R_g$ in polymer fluids\cite{prism}. 
%Two polymer chains that are separated by a distance smaller 
%than the range of their mutual interaction
%potential must influence one another's dynamics. 
Such
interactions on length scales 
shorter than or equal to $R_g$ leads to the Rouse equation's failure  
to adequately describe
short-time polymer dynamics.

The inconsistency of the  Rouse equation with computer simulation data 
of short-time
unentangled\cite{Paul,Paul1} and entangled polymer dynamics in the 
melt state, and in concentrated
solutions\cite{Kremer,Hall}, has been known 
for many years. 
In the short-time regime, $t \ll \tau_{Rouse}=2 R_g^2/(\pi^2 D_{Rouse})$,
the single-chain center-of-mass (c.o.m.) mean-square-displacement, 
$\Delta R^2(t)$, of unentangled 
polymers
shows anomalous diffusion.
For times $t  \ll \tau_{Rouse}$
$\Delta R^2(t)$ 
crosses over from the short-time ballistic dynamics to long-time Fickian diffusion
($\Delta R^2(t) \propto t$)
through an intermediate regime ($\Delta R^2 (t) \propto 
t^{0.8 \div 0.9}$)\cite{Paul,Paul1,Grest}.
Only at long times ($t \ge \tau_{Rouse}$) does the system obey 
Rouse dynamics and  
Fickian diffusion is
recovered.
Entangled
polymer fluids show the same behavior
in the short-time regime where Rouse dynamics
is expected to hold\cite{Kremer,Hall}.
The short-time anomalous dynamics
only appears in melts and
in solutions above the 
polymer overlap concentrations\cite{DoiEdw,Kremer,Hall}, further  
supporting the hypothesis that their short-time anomalous dynamics is
a consequence of intermolecular interactions. 

In a recent paper, we derived a Generalized Langevin Equation 
for the Cooperative Dynamics (CDGLE) of interacting, completely flexible 
polymer fluids\cite{CDGLE}, where explicitly intramolecular and intermolecular
forces are included. The CDGLE is here implemented to treat
the heterogeneous dynamics of unentangled 
semiflexible polymer melts with finite-size chains. The theory is found to be in excellent
agreement with computer simulation data performed by G.Grest and coworkers\cite{Grest}.

The GLE for a group of $n$ molecules undergoing slow cooperative dynamics
is derived from the 
first-principle 
Liouville equation by projecting
the dynamics of the entire
fluid 
onto the phase-space of the slow variables, i.e. the coordinates
of $\it{n}$ molecules
statistically comprised inside the volume
defined by the range of the mean-force potential,
$n=\rho R_g l^2 = \it{O}(\sqrt{N})$, with $\rho$ the monomer density.
An effective segment $a$ 
in molecule $i$
is driven by the intramolecular potential,  
$-\beta^{-1} ln \Psi({\bf r}^{(i)}(t))$ 
with 
${\bf r}^{(i)}(t)=\{ {\bf r}_1^{(i)}(t),{\bf r}_2^{(i)}(t), ..{\bf r}_a^{(i)}(t), ..
{\bf r}_N^{(i)}(t)\}$, 
the intermolecular time-dependent potential of mean-force,  
$-\beta^{-1} ln g[{\bf r}^{(j)}(t),{\bf r}^{(k)}(t)]$, 
and the random interactions with the 
surrounding fluid, where the projected random force is given by
${\bf F}_a^{Q(i)}(t)$. The segment  position 
${\bf r}_a^{(i)}(t)$ follows the
equation of motion\cite{CDGLE}:
\begin{eqnarray}
& & \zeta_{eff} \frac{d{\bf r}_a^{(i)}(t)}{dt}  =   
\frac{1}{\beta} \frac{\partial}
{\partial {\bf r}_a^{(i)}(t)}
\ln [\prod_{j=1}^n \Psi({\bf r}^{(j)}(t))  \nonumber \\
& &\prod_{k<j}^n g({\bf r}^{(j)}(t),{\bf
r}^{(k)}(t))] + {\bf F}_a^{Q(i)}(t) \ , \label{CDGLE}
\end{eqnarray}
with 
$\beta = (k_BT)^{-1}$.

The effective monomer friction coefficient, $\zeta_{eff}$, is given by the
linear combination of
the bare friction, $\zeta_0 = \beta <{\bf F}_a^i \cdot {\bf F}_a^{Q(i)}> / 3$,
conventionally assumed to be the Rouse friction,
and the intramolecular ($i$) and intermolecular ($j\neq i$)
memory functions containing time and space correlation of the
random forces:
\begin{eqnarray}
&& \zeta_{eff} = \zeta_0+
\frac{\beta}{3}\sum_{b\neq a}^{N} \int_0^\infty d t
<{\bf F}_a^{(i)}(0) \cdot {\bf F}_b^{Q(i)}(t)> \nonumber \\
&&
 + \frac{\beta}{3} \sum_{b=1}^{N}\sum_{j\neq i}^n \int_0^\infty d t
<{\bf F}_a^{(i)}(0) \cdot {\bf F}_b^{Q(j)}(t)>  \label{friction}
\ .
\end{eqnarray}
The coupling of the random forces acting on
different molecules is discarded in the traditional single-chain
approaches\cite{DoiEdw}, and it is negligible
for systems of uncorrelated molecules,
for which the pair distribution function
$g(r) \approx 1$.

With the purpose of describing the dynamics of finite-size semiflexible polymers,
Eq.(~\ref{CDGLE}) is reformulated in a matrix formalism, and semiflexibility
is introduced 
following the approach
developed by Bixon and Zwanzig at the simple
Rouse level\cite{BixonZ}.
The new equation of motion
applies to
the overdamped dynamics of a fluid of interacting
molecules of any kind, including semiflexible polymers of finite length.
It recovers the traditional single-molecule equation\cite{DoiEdw}
in the limit $\it{n}=1$, and/or
with the fluid described as a continuum, $g(r,t)=1$.
In the case $\it{n}=2$ and $N=1$, it
recovers the diffusive equation for binary interacting particles
\cite{2particles}. Thus our equation correctly reproduces the dynamics of systems 
characterized by dominant
intramolecular or intermolecular interactions.

Since 
Eq.(~\ref{CDGLE}) is obtained by projecting the fluid
dynamics onto the extended set of intra- and intermolecular
slow variables, the memory function contributions are minimized and 
can be neglected in first approximation\cite{hansen}. 
In addition, if only the intramolecular polymer coordinates are chosen as slow
variables, and the memory function is discarded, this procedure recovers
the traditional Rouse equation. 
The  memory function contribution to the intramolecular equation 
becomes relevant for melts of high molecular weight polymers, where it
describes the cross-over from unentangled to entangled polymer
dynamics\cite{pmc}.

The GLE is decoupled in center-of-mass diffusion and intramolecular
dynamics by a transformation of variables from beads to bond monomer 
coordinates\cite{marina}.
The solution of the c.o.m. diffusive equation requires the derivation of
an analytical expression 
for the c.o.m. intermolecular potential. 
We start from the four-point distribution function
which relates
a pair of monomers belonging to different polymers, and 
their c.o.m. positions. The distribution is factorized in the product of two-point 
distribution functions and integrated 
over the monomer coordinates
$g(R)\approx \int d{\bf r}_a \Psi(r_a^{(j)})
\int d{\bf r}_b \Psi(r_b^{(k)}) g(r_{a,b})$. 
In the special case of polymer molecules we adopt the thread-PRISM expression
of the monomer pair distribution
function\cite{prism}, and the Gaussian intramolecular distribution\cite{DoiEdw},
$\Psi(r_a)=[3/(2 \pi R_g^2)]^{3/2} exp[- 3 r_a^2/(2 R_g^2)]$.
The effective potential acting between the centers-of-mass of a pair of
molecules inside a region of 
slow cooperative dynamics is approximated by 
\begin{eqnarray}
& & w(R) = - \ln g(R) \approx \frac{27 \sqrt{2}}{4 \pi \sqrt{\pi}} \frac{1}
{\sqrt{N}\rho^*} \label{pot} \\
& & [1 - 108 \pi^{-2} (\rho^*)^{-2}(N)^{-1}] e^{-3 R^2/(4 R_g^2)} \ . \nonumber
\end{eqnarray}
$w(R)$ is Gaussian and finite for all 
distances, with a range on the order of the radius of gyration.
At full polymer-polymer
overlap, 
$w(0)$ decreases with increasing reduced fluid density,
$\rho^*=\rho l^3 $, reflecting the transition towards an incompressible
system. $w(0)$ also decreases with increasing
polymer molecular weight due to the increasing interpenetrability of the polymer
chains. 
Figure 1 shows a comparison between Eq.(~\ref{pot})
and the potential extracted from
the trajectories of molecular dynamics computer simulations
for melts of unentangled
polyethylene chains
\cite{Grest}.
The simulations
are performed at constant volume and constant temperature,
using the experimental
densities at atmospheric pressure. We analyze trajectories from unentangled polyethylene
samples with increasing
degree of polymerization, as reported in Table I
(the entanglement degree of polymerization $N_e=136$). 
The only fitting parameter in the quantitative comparison, is the potential at
complete polymer overlap, $w'(0)$, which is compared
in Table I to $w(0)$.
All the samples are in good agreement with Eq.(~\ref{pot}).
The value of $w'(0)$ agrees quantitatively with the simulations for the high
$N$ samples. In the low $N$ samples the presence of the fine structure
at short distance,
due to the
local monomer packing, doesn't allow us to make a quantitative
comparison.

The data show a decrease of $w(0)$ with increasing molecular weigth
or density, in a greement with Eq.(~\ref{pot}), and with a recent
study on dilute and semidilute polymer solutions\cite{Hansenp}.
Eq.(~\ref{pot}) represents an improvement on the 
mean-field formula derived by Flory and Krigbaum, which predicts 
the opposite trend\cite{Hansenp}.
However in solution both simulations and renormalization group 
calculations shows that the $N \rightarrow \infty$ scaling regime is finite.
In our study of polymer melt dynamics, the small range of
$N$ available does not allow us to investigate this feature.

From Eq.(~\ref{pot}) we derive the intermolecular force between the
c.o.m. of two polymers. In Eq.(~\ref{CDGLE}) this is moltiplied
by the number of interacting polymers
$G(t) R(t) =- \sqrt{N} \rho^* d W(t)/ dR(t)$.
The time evolution of the intermolecular center-of-mass potential 
follows the time-dependence of the
interpolymer
distance. 
With the purpose of obtaining an analytical solution of Eq.(~\ref{CDGLE}),
we approximate $R^2(t)$ with its statistical average over the polymer
configuration space, 
\begin{eqnarray}
& & G(t) R(t)
\approx \frac{81 \sqrt{2}}{8 \pi \sqrt{\pi}} R_g^{-2}
[1 - 108 \pi^{-2} (\rho^*)^{-2}(N)^{-1}] \nonumber \\
& & e^{-3 <R^2(t)>/(4 R_g^2)} R(t)
\label{gt} \ .
\end{eqnarray}
%In a multi-chain description the many-body distribution 
%function is approximated by the product of $n$ pair distribution functions.
%The relative intermolecular distance contains contributions from all the
%pairs of interacting molecules, $<R^2(t)>\approx n \Delta R^2(t)-
%6 D_{m.c.} t$, where $\Delta R^2(t)$ is the single
%chain mean-square displacement. The
%many-chain cooperative
%diffusion coefficient, $D_{m.c.}=(\beta \rho^* N\sqrt{N} \zeta_{0})^{-1}=D_{Rouse}/(
%\rho^* \sqrt{N})$,
%approaches zero in melts of high molecular weight polymer
%chains, at low temperature, and/or at high liquid density.
When Eq.(~\ref{gt}) is introduced in Eq.(~\ref{CDGLE}), the latter
reduces to 
a set of coupled Rouse equations with explicit 
intermolecular contributions
\begin{eqnarray}
\zeta_{0} \frac{d {\bf R}_{c.m.}^{(i)}(t)}{d t}& = &
G(t) [ {\bf R}_{c.m.}^{(i)}(t)
- {\bf R}_{t}(t)]
+ {\bf F}^{i}(t)  \ ,
\label{rai}
\end{eqnarray}
which obey the fluctuation-dissipation condition $<{\bf F}_{\alpha}^{i}(t) \cdot
{\bf F}_{\gamma}^{j}(t')> = 6 \zeta_0 \beta^{-1} \delta_{\alpha \gamma}
\delta_{ij}
\delta(t-t')$,
where $\alpha,\gamma=x,y,z$.
${\bf R}_{t}(t)=n^{-1}\sum_{k=1}^n{\bf R}_{cm}^{(k)}(t) $ is the coordinate
of the c.o.m. of the dynamical aggregate.

The advantage of adopting the approximate form of the intermolecular force
is that the GLE becomes solvable using the standard techniques of transformation
to normal mode coordinates. The set of coupled equations, Eq.(~\ref{rai}),
reduces to two
GLEs in the relative and cooperative variables, from which the 
relative many-chain intermolecular distance 
$<R^2(t)>\approx n \Delta R^2(t)-
6 D_{m.c.} t$, where $\Delta R^2(t)$ is the single
chain mean-square displacement. The
many-chain cooperative
diffusion coefficient, $D_{m.c.}=(\beta \rho^* N\sqrt{N} \zeta_{0})^{-1}=D_{Rouse}/(
\rho^* \sqrt{N})$,
approaches zero in melts of high molecular weight polymer
chains, at low temperature, and/or at high liquid density.
Since the intermolecular force depends at each instant on the value of the
intermolecular distance $<R^2(t)>$, we solve Eq.(~\ref{rai}) self-consistently
for small time increments ($\Delta t =10^{-2} \ ps$).
By this procedure we take into account the 
change in the intermolecular
interaction with the dynamical evolution of the system.

Using Eq.(~\ref{rai}), we calculate the c.o.m. mean-square
displacement, $\Delta R^2(t)=<(R_{c.m.}(t)-R_{c.m.}(0))^2>$.
Since the appearance of anomalous c.o.m. dynamics is 
related to the amplitude of $G(t)$ (when $G(t) \rightarrow 0$ single chain 
uncorrelated dynamics is recovered)
the anomalies are confined in the time scale $t \leq \tau_{Rouse}$,
which is the time necessary for the system to travel outside the
range of the mean-force potential ($R^2(t) \geq R_g^2$).

%In a few limiting
%cases, the single-chain mean-square displacement, $\Delta R^2(t)$, 
%can be reduced to a simple analytical form.
%For strongly interacting polymers 
%the theory predicts subdiffusive dynamics for 
%$t \geq \tau_1$ with $\tau_1=\zeta_{0} /(2 G(0))$. 
%The physical explanation of this anomalous behavior is that
%the molecule is temporarily trapped due to the intermolecular potential of
%the neighboring molecules, and anomalous slowing down of
%the dynamics takes place.
%If we assume that the infinitesimal translation of the c.o.m. is
%linear in time, $\Delta R^2(t)\approx 6 D t$, we obtain an analytical equation for 
%the self-consistent expression of $\Delta R^2(t)$.
%For high molecular weight, slowly relaxing polymer melts ($9 D t << 2 R_g^2$) 
%the dynamics crosses over from subdiffusive to Fickian diffusion
%at $t \approx 2 R_g^2/(9 D)$. The anomalous subdiffusive behavior
%doesn't occur in melts of small, highly
%mobile molecules.  

We test the validity of our approach by comparing the c.o.m. 
mean-square displacement calculated from the solution of Eq.(~\ref{rai}), 
with the data 
from computer simulations for the samples previously analyzed.
The values of $\rho$, $T$, $N$, $l$ and $\zeta_0$ are the input to the
equation together with the effective value 
of the potential at initial time and
zero intermolecular distance, $w'(0)$, calculated from the fitting of
the potential. Here the statistical segment is much larger than the chemical 
bond because of the intrinsic stiffness of the polyethylene molecule.
Figures 2 and 3 show that the diffusive equation with the soft-core 
intermolecular potential
reproduces quantitatively the anomalous subdiffusive
c.o.m. dynamics for all of the samples investigated, and in the 
complete range of time scales.
The slope of the subdiffusive c.o.m. regime is reported in Table I, and
it is found to decrease with increasing degree of polymerization. 
The degree of dynamical anomaly
increases with increasing $N$ and decreasing $T$, as the system moves towards
its glass transition. For the range of $N$ investigated in this study
a scaling regime has yet to be reached.

To describe the
polymer flexibility we adopt
a freely rotating chain model for segments of length
$l=1.53 \stackrel{\circ}{A}$
with an effective
stiffness
parameter $g=0.74$
($g=<{\bf l}_i \cdot {\bf l}_{i+1}> / l^{2}$, ${\bf l}_i={\bf r}_{i+1}
- {\bf r}_i$). The value of $g$ is obtained from the polymer mean-square
end-to-end distance calculated from the simulation data,
and reproduces the experimental characteristic ratio for a finite
polyethylene chain.
The introduction of the explicit description of the local stiffness does not affect
the dynamics at the c.o.m. level. This is because both the flexible model with 
renormalized segments and the the explicit semiflexible description are
devised to reproduce the global chain properties.
In figures 2 and 3 the c.o.m. diffusion is identical if we use a semiflexible or a 
renormalized flexible model.

The monomer dynamics depends strongly on the local chemical
structure, and chain stiffness. However, at
the monomer level, the intermolecular contributions  
are negligible. Our mode analysis shows that only the lowest index, global
modes are affected by the intermolecular potential in agreement 
with previous studies\cite{Paul,Kremer,modes}.
In analogy with the static picture,
we argue that the insensitivity of the local dynamics to the
intermolecular forces is
due to the presence of similar and
compensating monomer-monomer
intramolecular and intermolecular excluded-volume
interactions\cite{DeGennes}.
In the regime of $t \leq \tau_{Rouse}$, the data show a dynamics
slightly faster than the Rouse prediction of $\Delta r^2(t) \propto t^{0.5}$.
This effect is predicted by our diffusion equation with included local polymer 
semiflexibility. 
Figure 3 shows that
the monomer mean-square displacement calculated with the semiflexible 
approach is in better agreement with
the simulations than the Rouse function. The inclusion, or exclusion, of the
intermolecular contribution, does not modify the monomer diffusion. 

In conclusion we show that the center-of-mass intermolecular
potential in polymer melts 
can be described at a good level of approximation by an effective
Gaussian potential. The introduction of the local 
intermolecular forces in the GLE strongly improves the description of the
global dynamical properties of polymer melts in the short-time domain. The
intermolecular forces, however, do not affect the dynamics on the local
monomer scale, which is instead strongly dependent on 
the local polymer semiflexibility.

We are grateful to G.S.Grest for sharing the computer simulation
trajectories.
Acknowledgment is made to the donors of The Petroleum Research Fund, 
administrated by the
ACS, for partial support of this research.
We also acknowledge the support of the National Science Foundation under 
the grant DMR-9971687.

\end{multicols} 

\begin{tabular}{|c|c|c|c|c|c|c|} \hline \hline
Polymer & $N$ & $T \ [K]$ & $\rho \ [g/cm^3] $ & 
$l [\stackrel{\circ}{A}]$ & $\frac{w(0)}{w'(0)}$ 
& $slope $ \\
\hline
$C_{10}H_{22}$ & $10$ & $298$ & $0.7250$ & 3.26 & 2 & 0.97 \\
$C_{16}H_{34}$ & $16$ & $298$ & $0.7703$ & 3.84 & .4 & 0.94 \\
$C_{16}H_{34}$ & $16$ & $323$ & $0.7531$ & 3.75 & .6 & 0.96 \\
$C_{16}H_{34}$ & $16$ & $373$ & $0.7187$ & 3.66 & .8 & 0.97 \\
$C_{30}H_{62}$ & $30$ & $400$ & $0.7421$ & 4.02 & .3 & 0.90 \\
$C_{44}H_{62}$ & $44$ & $400$ & $0.7570$ & 4.18 & .3 & 0.85 \\
\hline \hline
\end{tabular}
\section*{TABLE CAPTION}
Table I: Simulation parameters, and fitting parameters.

\section*{FIGURE CAPTIONS.}
FIG. 1: Comparison between the analytical expression of the 
soft-core center-of-mass Gaussian potential, Eq.(~\ref{pot}), 
and computer simulations data, vs. the c.o.m. interpolymer distance
normalized by the polymer end-to-end distance, $R_{ete}=\sqrt{N} l$.

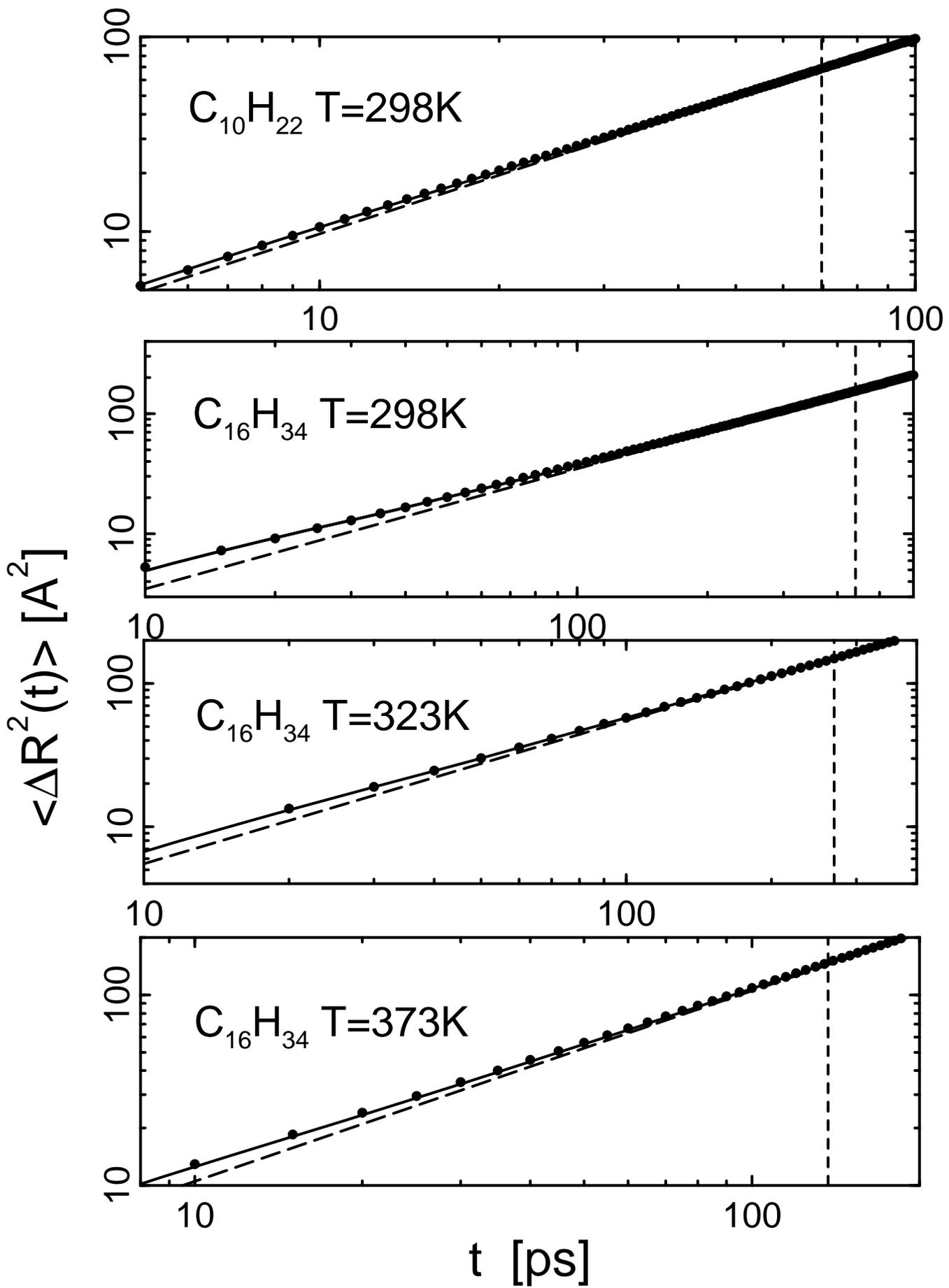
FIG. 2: Center-of-mass mean-square displacement as a function of time. Best fit of
the molecular dynamics simulation data (filled circle) with the Rouse equation (dashed
line), and with the intermolecular diffusion equation, Eq.(~\ref{rai}), for polymer
melts
(full line). The short-dashed lines indicate
the longest Rouse relaxation time, $\tau_{Rouse}$.

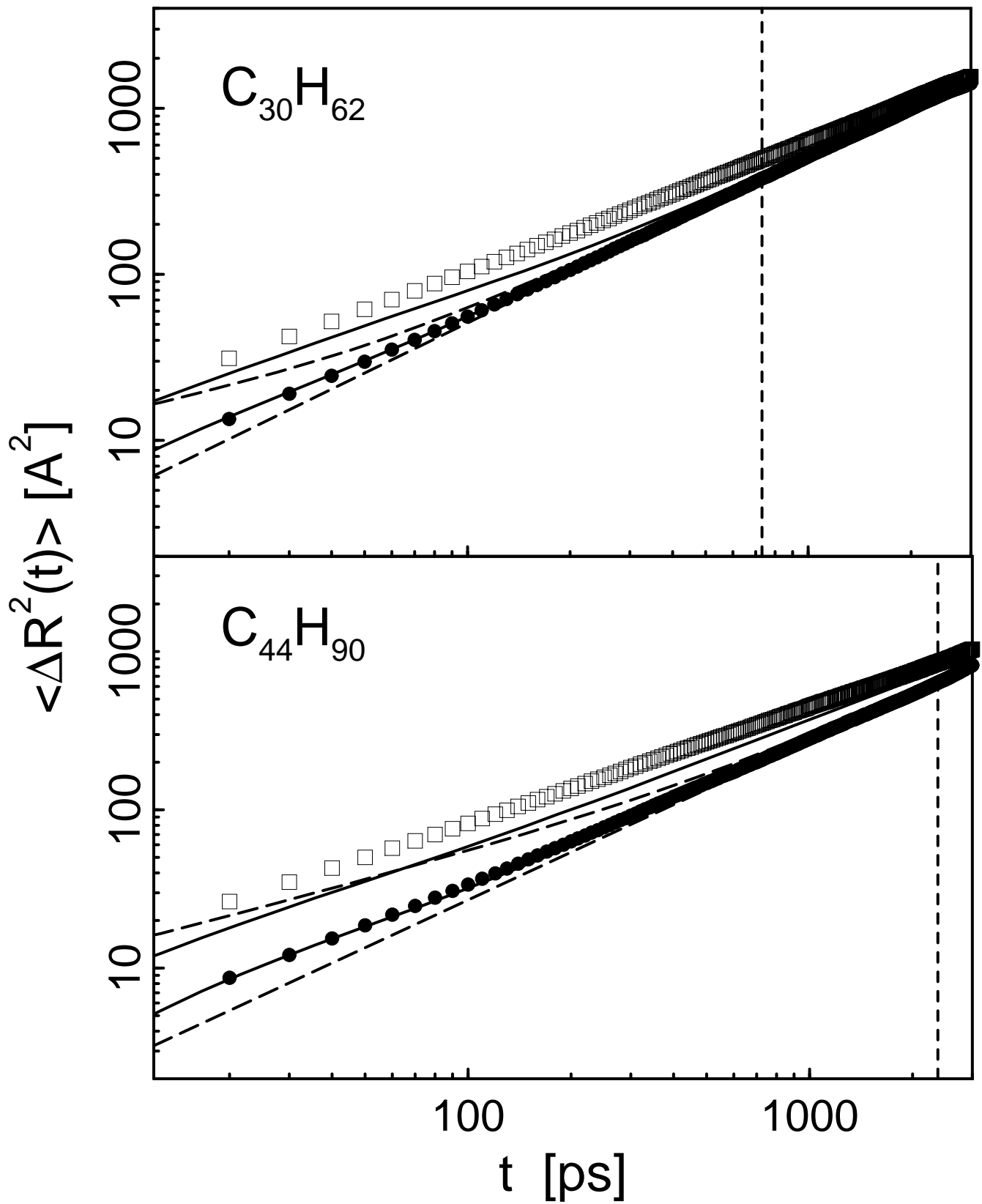
FIG. 3: Monomer and center-of-mass mean-square displacements as a 
function of time. Best fit
of the molecular dynamics simulation data (filled circle for c.o.m. and open squares for
monomer dynamics) with the Rouse equation (dashed 
line), and with the diffusion equations for semiflexible polymer
melts (full line).
The short-dashed lines indicate
the longest Rouse relaxation time, $\tau_{Rouse}$.

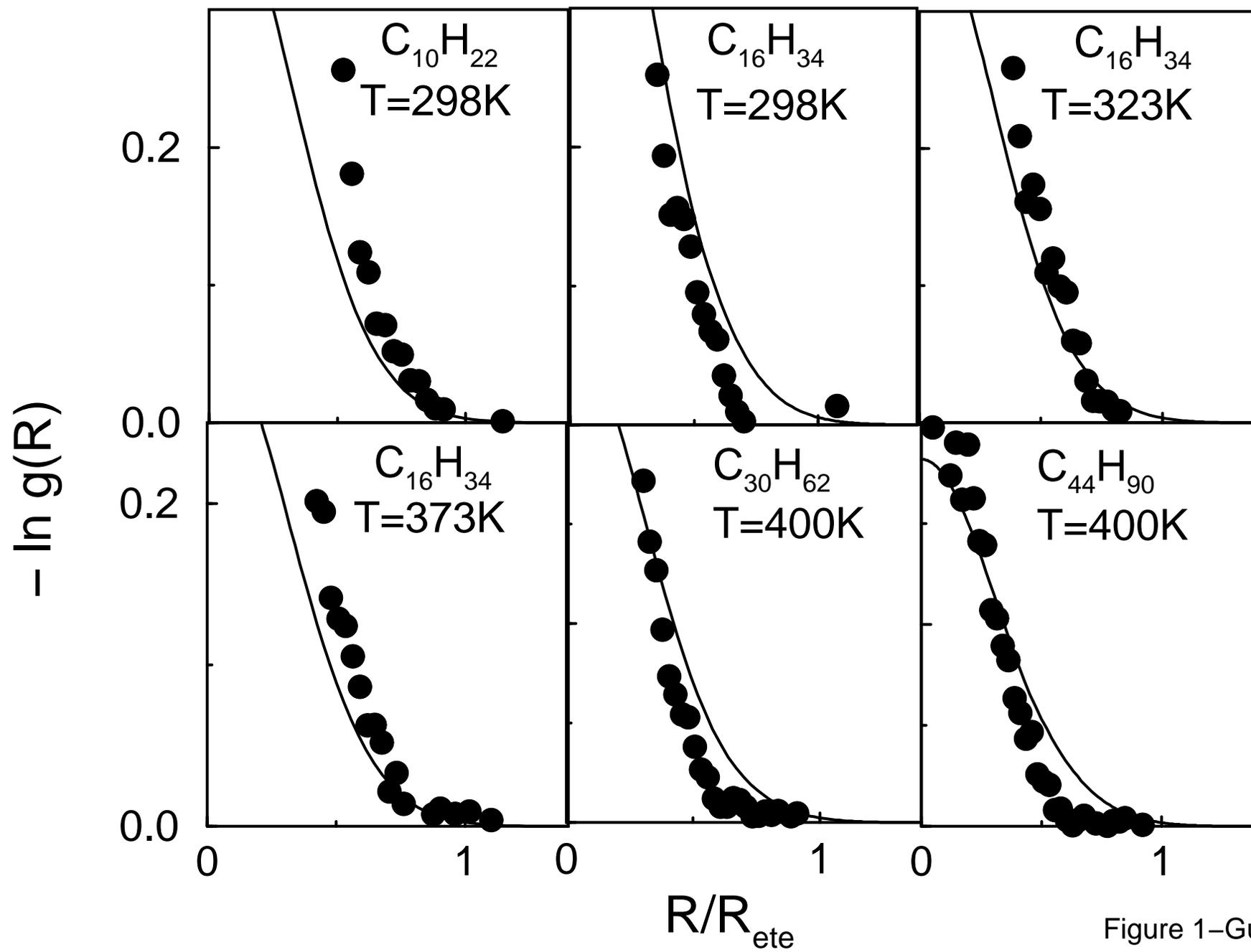
\end{document}